\documentclass[sigconf,screen]{acmart}

\usepackage[english]{babel}
\usepackage{blindtext}
\usepackage{float}
\usepackage{caption}
\usepackage{subcaption}
\usepackage{xcolor}
\usepackage{booktabs}
\usepackage{multirow}
\usepackage{hyperref}
\usepackage{siunitx}
\renewcommand{\sectionautorefname}{\S\kern-0.25em}
\renewcommand{\subsectionautorefname}{\S\kern-0.25em}
\usepackage{enumitem}
\usepackage[ruled,vlined]{algorithm2e}
\usepackage{placeins}
 
\renewcommand\footnotetextcopyrightpermission[1]{}

\acmDOI{}

\acmISBN{}

\acmPrice{}

\newcommand{\sysname}{DualPath\xspace}

\newcommand{\parabf}[1]{\medskip\noindent\textbf{#1}}
\newcommand{\paraf}[1]{\noindent\textbf{#1}}

\begin{document}
\title{\sysname: Breaking the Storage Bandwidth \\ Bottleneck in Agentic LLM Inference}

\author{Yongtong Wu$^{1,3}$ \enspace Shaoyuan Chen$^{2,3}$ \enspace Yinmin Zhong$^{1,3}$ \enspace Rilin Huang$^1$\\ 
Yixuan Tan$^3$ \enspace Wentao Zhang$^3$ \enspace Liyue Zhang$^3$ \enspace Shangyan Zhou$^3$ \enspace Yuxuan Liu$^3$ \enspace Shunfeng Zhou$^3$ \enspace Mingxing Zhang$^2$ \enspace Xin Jin$^1$ \enspace Panpan Huang$^3$}

\affiliation{
\institution{
$^1$School of Computer Science, Peking University \quad
$^2$Tsinghua University \quad
$^3$DeepSeek-AI \quad
}
}

\renewcommand{\shortauthors}{Wu et al.}
\def\shorttitle{\sysname}

\begin{abstract}

The performance of multi-turn, agentic LLM inference is increasingly dominated by KV-Cache storage I/O rather than computation. In prevalent disaggregated architectures, loading the massive KV-Cache from external storage creates a fundamental imbalance: storage NICs on prefill engines become bandwidth-saturated, while those on decoding engines remain idle. This asymmetry severely constrains overall system throughput.

We present \sysname, an inference system that breaks this bottleneck by introducing dual-path KV-Cache loading. Beyond the traditional storage-to-prefill path, \sysname enables a novel storage-to-decode path, in which the KV-Cache is loaded into decoding engines and then efficiently transferred to prefill engines via RDMA over the compute network. \sysname combines this optimized data path --- which inherently avoids network congestion and avoids interference with latency-critical model execution communications --- with a global scheduler that dynamically balances load across prefill and decode engines.

Our evaluation on three models with realistic agentic workloads demonstrates that \sysname improves offline inference throughput by up to 1.87$\times$ on our in-house inference system. It can also improve online serving throughput by an average factor of 1.96$\times$ without violating SLO.

\end{abstract}

\maketitle

\section{Introduction}

Large Language Models (LLMs) are rapidly evolving from single-turn chatbots~\cite{chatgpt,DeepSeekV32} and standalone reasoners~\cite{chatgpt} into \emph{agentic systems} that can autonomously plan, invoke tools, and solve real-world tasks through \emph{multi-turn interactions}~\cite{chowa2026language,wang2024survey,xi2025rise,jiang2024survey,mohammadi2025evaluation}. In such settings, an LLM no longer serves isolated prompts; instead, it participates in long-running sessions where context accumulates over time~\cite{lin2025efficientagentscodesigninference}. As agentic applications become increasingly prevalent, multi-turn LLM inference has emerged as a critical workload in production systems, ranging from coding assistants~\cite{yang2024swe,wu2024autogen} to autonomous task agents~\cite{zhou2023webarena,li2024personalllmagentsinsights}.

This paradigm shift in applications has driven a significant transformation in LLM inference workloads: from traditional human-LLM interaction to human-LLM-environment interaction, called the \emph{agentic paradigm}. The typical pattern of human-model interaction involves users providing input, engaging in a few rounds of interaction with the LLM, and consuming the results generated by the LLM. By contrast, an agentic LLM may interact with an external environment, through tools such as a web browser and Python interpreter, over dozens or even hundreds of turns. Although each individual tool call or feedback is short (often hundreds of tokens), the context accumulates across turns and can grow to extreme lengths. As a result, agentic workloads become highly I/O-bound: the multi-turn, short-append pattern leads to very high KV-Cache hit rates --- typically $\geq 95\%$ \cite{chen2026concurhighthroughputagenticbatch} --- making the efficiency of KV-Cache loading, rather than pure computation, the dominant performance factor.

\begin{figure}[t!]
  \centering
  \includegraphics[width=\linewidth]{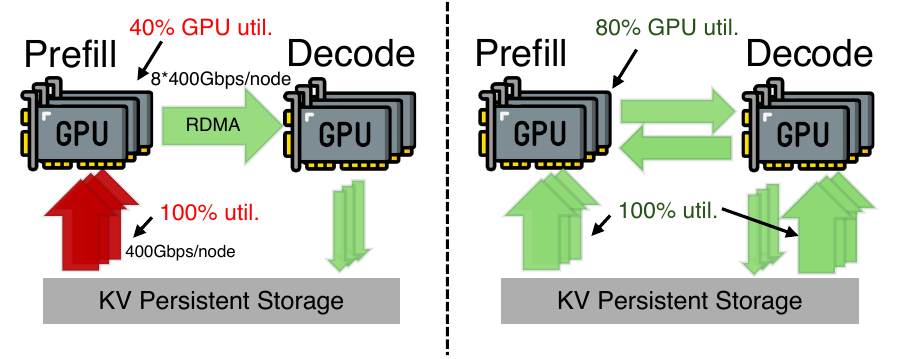}
  \vspace{-0.25in}
  \caption{\normalfont Existing bottleneck (left) and \sysname (right).}
  \vspace{-0.2in}
  \label{fig:teaser}
\end{figure}

To improve throughput under agentic workloads, existing LLM inference systems have converged on a common set of architectural patterns: \emph{layer-wise prefill}~\cite{LayerKV,SOSP25:PrefillOnly}, \emph{prefill--decode (PD) disaggregation}~\cite{OSDI24:DistServe,ISCA24:Splitwise,zhao2025insights}, and \emph{external KV-Cache storage}~\cite{ATC24:CachedAttention, LMCache, FAST25:Mooncake}.
In these systems, prefill engines load the KV-Cache in a layer-wise manner to accommodate as many requests as possible within a single batch. When prefill completes, decoding engines typically receive KV-Cache from prefill engines via a high-performance RDMA network. The decoding engines then generate tokens and store their KV-Cache in distributed storage to enable reuse across turns. %, instead of loading from storage, to avoid read amplification. % This separation allows better specialization and scalability, and has been widely adopted in recent systems.

However, this architecture also introduces a critical limitation. As shown in~\autoref{fig:teaser}, prefill engines must load large volumes of KV-Cache from remote storage.
As a result, \emph{prefill-side storage network bandwidth} becomes the throughput bottleneck of the entire system, even though decoding engines often have substantial unused storage network bandwidth.

This imbalance reveals a fundamental inefficiency in existing designs: storage network bandwidth is unevenly utilized across engines. The bandwidth of prefill engines are persistently saturated, while decoding engines remain underutilized. 
Simply provisioning more bandwidth to prefill engines is costly and often impractical in general-purpose clusters. Therefore, it is promising to exploit and combine the available I/O bandwidth of all engines, rather than overloading prefill engines alone, to accelerate KV-Cache loading for agentic LLM workloads.

Prior studies have attempted to alleviate the KV-Cache loading bottleneck.
Mooncake \cite{FAST25:Mooncake} caches KV-Cache in a distributed DRAM pool and employs an affinity-aware scheduler to maximize the DRAM KV-Cache hit rate.
However, it cannot be used in memory-constrained scenarios, such as the rollout phase in RL, where DRAM is occupied to hold large training state that is offloaded from HBM.
It is also not cost-effective in scenarios with enormous working sets (e.g., online serving), considering the cost comparison between DRAM and SSD.
Other attempts reduce the amount of KV-Cache data to retrieve \cite{EuroSys25:HCache} and reduce the retrieval overhead \cite{APSys25:TARDIS, SC25:Phoenix}.
However, they do not solve the inherent inefficiency caused by storage I/O imbalance between different engines.

In this paper, we present \sysname, a new LLM inference system that rethinks KV-Cache loading in modern inference architectures for agentic workloads. The key insight behind \sysname is that KV-Cache loading does not have to be prefill-centric. While existing systems always load KV-Cache directly from storage into prefill engines, they cannot utilize the remote storage bandwidth of decoding engines. \sysname leverages this observation by enabling \textbf{dual-path KV-Cache loading}: in addition to the conventional storage-to-prefill path, KV-Cache can be loaded into decoding engines and then transferred to prefill engines via high-performance RDMA. By dynamically selecting between these paths, \sysname redistributes network load and alleviates prefill-side bandwidth pressure.

Realizing this design raises two challenges.
First, introducing an extra loading path introduces complex traffic patterns and potential interference with collective primitives in model execution, which can degrade overall performance if unmanaged.
Second, the system must decide online which loading path to use under dynamic and heterogeneous workloads, and ensure load balance across both GPUs and NICs simultaneously.
To address these challenges, \sysname adopts (1) an optimized dual-path loading data path design, which introduces no inherent congestion under common P/D ratios, (2) a NIC-centric traffic management approach to isolate KV-Cache traffic from latency-sensitive model inference communications, and (3) a dynamic scheduling policy that jointly balances computation and network utilization across prefill and decoding engines.

We implement \sysname on top of a modern inference stack and evaluate it using representative agentic workloads with long contexts and high cache reuse.
Experiments show that \sysname significantly improves system throughput and the first token latency, while maintaining the latency between tokens. In agentic inference scenarios, \sysname increases end-to-end throughput by up to 1.87$\times$ for offline inference, and improves the online serving throughput by 1.96$\times$ on average.

In summary, this paper makes three contributions:
\begin{itemize}[leftmargin=*]
  \item We identify the I/O-bound nature of multi-turn, agentic LLM workloads and show that KV-Cache loading dominates system performance under modern LLM inference architectures.
  \item We present \sysname, an inference system that introduces dual-path KV-Cache loading and leverages decoding-engine bandwidth to resolve prefill-side bottlenecks.
  \item We design and evaluate a workload-aware scheduling algorithm that dynamically balances computation and network resources, significantly improving balance on realistic workloads.
\end{itemize}

\section{Background}

\label{sec:background}

\subsection{LLM Inference Preliminary}
LLM inference is becoming one of the most important system workloads recently.
Popular LLMs utilize decoder-only transformer architecture, comprising stacked blocks with attention layers and feed-forward networks (FFNs).
Attention layers enable token interactions within requests, while FFNs process tokens independently.
The model predicts subsequent tokens based on preceding ones, storing attention keys and values as \emph{KV-Cache} in HBM to avoid recomputations.

\parabf{PD-disaggregated Inference.} \emph{Prefill–decode (PD) disaggregation}  \cite{OSDI24:DistServe, ISCA24:Splitwise} separates the prefill phase from the decode phase, assigning them to dedicated prefill engines (PEs) and decode engines (DEs), respectively. The two phases exhibit distinct compute and memory patterns: prefill is compute-intensive and batched, while decode is memory-bound and latency-sensitive. With PD disaggregation, PEs load the hit KV-Cache and perform prefilling; then, they transfer the KV-Cache to DEs, which perform autoregressive decoding. This design reduces interference between phases, enables stage-specific optimizations, and improves scalability, making it the de facto architecture for modern LLM serving. To support multi-turn conversations, the KV-Cache is often stored in distributed storage for reuse across turns.

\parabf{Layerwise Prefill.} Long-context prefill is bottlenecked by HBM capacity, as both activations and the KV-Cache for the entire batch must reside within it, forcing limited batch sizes and leading to poor GPU utilization. LayerKV \cite{LayerKV} and PrefillOnly \cite{SOSP25:PrefillOnly} address this problem by exploiting the strong locality in prefill computations: each layer requires only its own layer-specific KV-Cache. Consequently, the KV-Cache can be allocated and freed per layer, and the GPU holds only one layer's KV-Cache for the forward batch. This increases the effective batch size (in tokens) by approximately a factor equal to the number of layers, boosting prefill throughput.

\subsection{Agentic Use of LLMs}

\begin{figure}[t!]
  \centering
  \includegraphics[width=\linewidth]{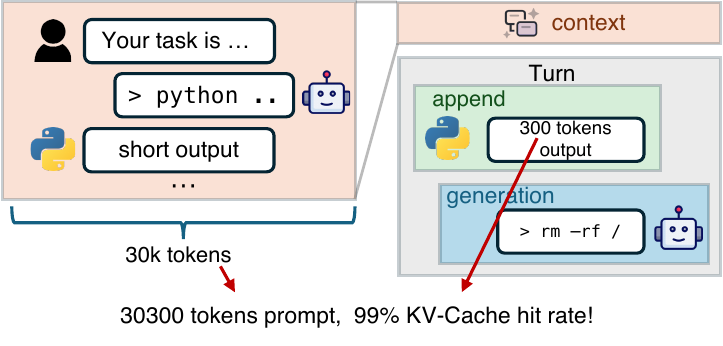}
  \vspace{-0.3in}
  \caption{\normalfont Agent trajectory example.}
  % \vspace{-0.2in}
  \label{fig:workload}
\end{figure}

LLMs increasingly power \emph{agentic} applications that perform multi-turn reasoning and interact with an environment (via e.g., terminal commands, code execution, or asking for human feedback) over long sessions. 
As shown in \autoref{fig:workload}, in a typical \emph{turn}, the model receives a prompt composed by the previous \emph{context} plus some newly \emph{appended tokens} (often tool output or user input) and \emph{generates} the next action or response. A single agent run is a \emph{trajectory} of dozens or even hundreds of turns: the context grows turn-by-turn and can reach up to one million tokens \cite{claudeopus46, googlegemini3pro}. 
Because most of the context, typically >95\% tokens in our traces, is reused across rounds, the vast majority of tokens in each round can hit the KV-Cache; only the newly appended context needs prefill computation. Due to the extreme length of agent trajectories, DRAM and HBM-based KV-Cache storage like Mooncake \cite{FAST25:Mooncake} can only store a small proportion of KV-Caches, necessitating the use of larger yet cheaper external SSD-based KV-Cache storage \cite{3FS}.

The agentic LLM inference workload is also prevalent in agent LLM training, which often adopts \emph{reinforcement learning} (RL) approaches. In a typical RL training loop, the agent LLM first undergoes a \emph{rollout} phase, where it is prompted to generate a large number of multi-step agent trajectories. These trajectories are then scored by a separate reward model. Finally, the LLM parameters are updated to increase the likelihood of high-scoring outputs and reduce the likelihood of low-scoring ones. During the rollout phase, substantial data (like reward model and optimizer states) is offloaded to host DRAM, further constraining the available DRAM for KV-Cache. This reinforces the need for external, high-capacity KV-Cache storage that can accommodate long agentic rollout contexts efficiently.

\subsection{Modern AI Data Center Architecture}

Modern AI data centers are purpose-built logical supercomputers engineered to handle large-scale generative AI training and inference workloads. For example, in a standard NVIDIA DGX SuperPOD \cite{nvidia2023superpod}, each node is equipped with 8 Hopper GPUs interconnected via high-speed NVLink. Each GPU is paired with a dedicated 400 Gbps compute NIC (\emph{CNIC}, also known as east-west NIC), which maximizes inter-node communication bandwidth. Independent of the compute fabric, each node also features a storage NIC (\emph{SNIC}, also known as south-north NIC) up to 400 Gbps, providing fast access to datasets, model checkpoints, and on-disk KV cache.

A fundamental principle of this architecture is that the compute network and the storage network are isolated from each other \cite{zhao2025insights}. This separation is essential to maximize both storage and application performance. By isolating high-intensity east-west compute traffic between GPUs from storage traffic, the architecture prevents interference between them, and drastically reduces compute communication latency. This design also ensures that the inter-GPU communication remains highly reliable and predictable even when performing data-intensive tasks such as reading large datasets or writing multi-terabyte model checkpoints.
\section{Bottleneck \& Motivation}
\label{sec:motivations}

We observe severe GPU underutilization during agentic inference tasks.
Our investigation reveals that KV-Cache loading speed is the bottleneck due to the limited bandwidth of the single storage NIC on each node.
Analysis demonstrates that three decisive factors jointly cause this bottleneck, as discussed below.

First, agentic workloads exhibit high KV-Cache hit rates, which \emph{require more I/O and less computation}, thus creating a severe I/O bottleneck.
Agentic workloads are naturally long-context, short-append, and multi-turn.
On each turn, the GPU needs to read the KV-Cache of the entire context from persistent storage and perform prefill computation for appended tokens.
Our trace collected from representative coding tasks shows the mean number of rounds is 157, demonstrating the tendency of LLMs to engage in multi-turn interactions. The average context length is 32.7k, while the append length mean is only 429, which means a KV-Cache hit rate of 98.7\%. 
In such a scenario, the cache-compute ratio, defined as the ratio of KV-Cache to load and the computation needed, is approximately 22 GB/PFLOP for DeepSeek-V3.2 \cite{DeepSeekV32}, posing a significant bottleneck on storage bandwidth. 
Note that the KV-Cache size of DeepSeek MLA model is already highly optimized; for models with larger KV-Cache sizes (see \autoref{tab:cache-compute-ratio}), the situation is even worse. 
The ratio of DeepSeek-V3.2 is higher than DeepSeek-V3 \cite{deepseekv3}, benefiting from its sparse attention design, lowering computation demands.

\begin{table}[t]
  \centering
  \caption{\normalfont Cache-compute ratio with append length 429, across context lengths (16k--64k). KV-Cache data type defaults to FP8 unless specified.}
  \begin{tabular}{lc}
  \toprule
  Model & GB/PFLOP (16K--64K) \\
  \midrule
  % Gemma3-27B~\cite{gemmateam2025gemma3technicalreport} (FP16) & 1.0 \\
  Qwen2.5-32B~\cite{qwen2025qwen25technicalreport} (FP16) & 117-267 \\
  % DeepSeek-V3.2 27B~\cite{DeepSeekV32} & 70--184 \\
  GPT-OSS-120B~\cite{GPT-OSS} & 47--95 \\
  Qwen3-235B-A22B~\cite{yang2025qwen3technicalreport} & 39--60 \\
  DeepSeek-V3.2 660B~\cite{DeepSeekV32} & 13--36 \\
  % Kimi-K2-Base 1T~\cite{kimiteam2026kimik2openagentic} & 0.028 \\
  DeepSeek-V3 660B~\cite{deepseekv3} & 4.8--5.8 \\
  \bottomrule
  \end{tabular}
  \label{tab:cache-compute-ratio}
\end{table}

\begin{figure}[t!]
  \centering
  \includegraphics[width=1.05\linewidth]{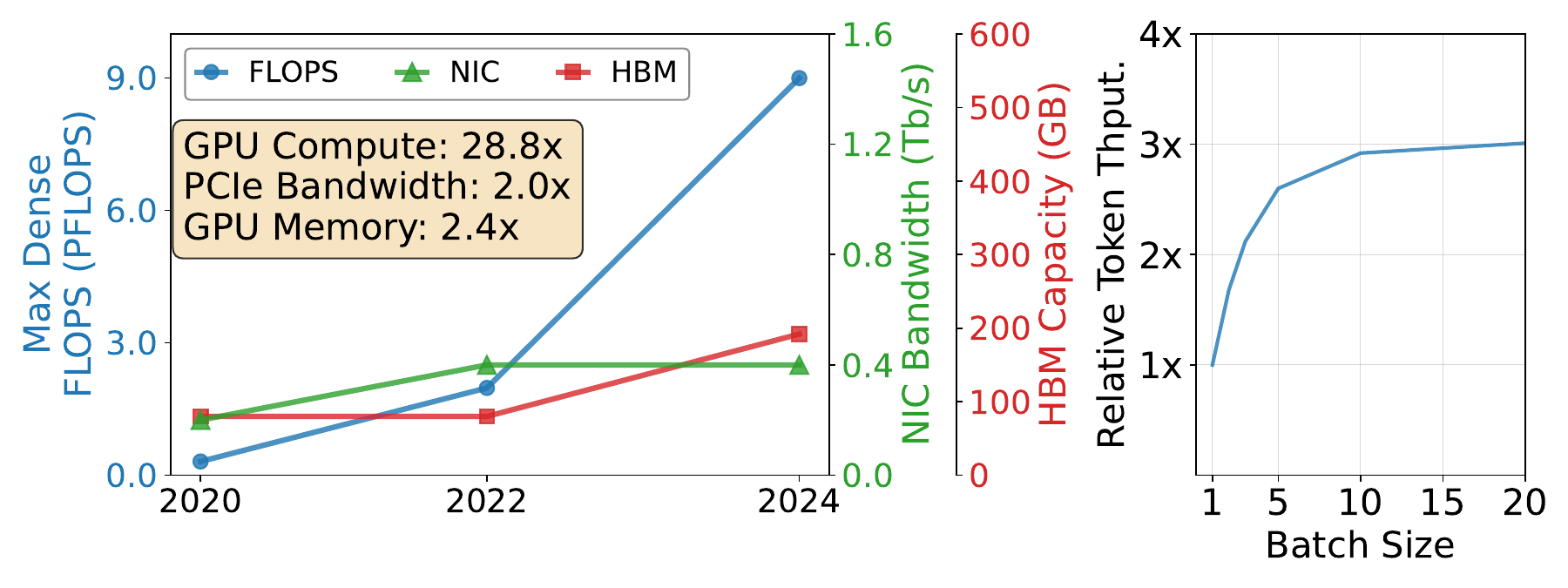}
  \vspace{-0.35in}
  \caption{\normalfont Left: Hardware trends of NVIDIA GPUs. Right: Relative token throughput with varying request batch size (each request has 30K context with 300 tokens appended).}
  \vspace{-0.2in}
  \label{fig:motivation}
\end{figure}

Second, the \emph{hardware evolution trend} is not well suited for agentic inference workloads.
In recent years, network bandwidth and HBM capacity have lagged behind the growth of GPU FLOPS, which drives us to run into memory and communication walls under agentic workloads.
As shown in \autoref{fig:motivation}, from NVIDIA Ampere to Blackwell, the I/O-compute ratio decreases by 14.4$\times$.
Low NIC bandwidth limits KV-Cache loading speed, making GPUs idle.
In addition, small HBM capacity limits the token batch size for GPU kernels \cite{ICLR24:FlashAttention2, FlashInfer, DeepGEMM, FlashMLA} to compute at the same time, hindering full utilization of compute units such as Tensor Core \cite{SOSP25:PrefillOnly}. % Small HBM capacity also limits the overlapping between computation and KV-cache transfers, which requires extra buffer, causing transfer latency to fall on the critical path.

Third, existing LLM inference systems exhibit severe \emph{storage network utilization imbalance} across different engine types. In prevalent PD-disaggregated systems, the KV-Cache for hit tokens is loaded exclusively by prefill engines directly from remote storage. This design centralizes all storage I/O pressure on the prefill-side SNICs, while the SNICs on decode engines remain largely idle. Consequently, the aggregate storage network bandwidth cannot be fully harnessed. % This static assignment of the KV-Cache loading role creates a fundamental and inefficient imbalance in storage network bandwidth utilization across engine types.

The above analysis demonstrates that the fundamental performance issue for agentic inference on PD-disaggregated architecture is the high I/O demand for KV-Cache retrieval and unbalanced storage network bandwidth utilization across inference engines. Meanwhile, we observe that the network traffic of the compute network, which has much larger aggregate bandwidth than the storage network, exhibits an intermittent pattern: collective operations used in model inference burst in sub-millisecond intervals. Therefore, an opportunity naturally emerges: we can utilize the SNIC bandwidth of decode nodes to load KV-Cache from storage, and transfer it back to the prefill nodes, utilizing the spare bandwidth of the faster compute network.

\section{\sysname System Overview}
\label{sec:system-overview}

To break the prefill-side storage I/O bottleneck, we propose a \textbf{dual-path loading} architecture that fundamentally rethinks how KV-Cache is retrieved in PD-disaggregated inference. Based on this architecture, we design and implement \sysname.
\sysname adopts two widely-adopted techniques demonstrated in \autoref{sec:background}:
(1) \textbf{PD Disaggregation}~\cite{OSDI24:DistServe,ISCA24:Splitwise}, which separates prompt and decode processing for better efficiency.
(2) \textbf{Layerwise prefill}, which avoids HBM bottlenecks recognized by LayerKV~\cite{LayerKV} and PrefillOnly~\cite{SOSP25:PrefillOnly} on prefill engines and improves GPU utilization.

\begin{figure*}[t!]
  \centering
   \begin{subfigure}[b]{0.45\textwidth}
     \centering
      \includegraphics[width=\textwidth]{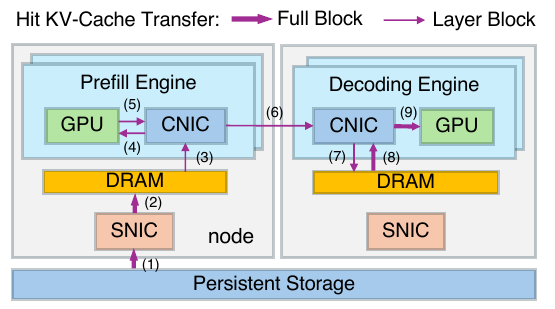}
      \caption{\normalfont PE Read Path}
      \label{fig:pe_flow}
    \end{subfigure}
    \hspace{6mm}
    \begin{subfigure}[b]{0.45\textwidth}
        \centering
      \includegraphics[width=\textwidth]{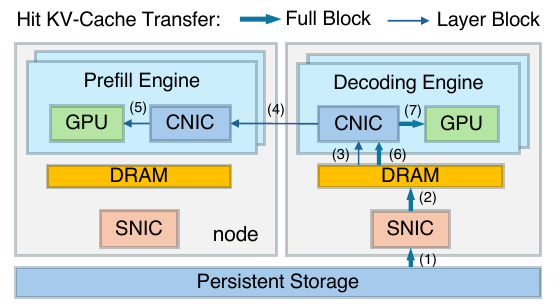}
      \caption{\normalfont DE Read Path}
      \label{fig:de_flow}  
    \end{subfigure}
    \caption{\normalfont Dual-path loading illustration. The scheduler dynamically distributes data traffic between the two paths.}
    \label{fig:dual-path-loading}
\end{figure*}

Our system consists of the following components:
\begin{itemize}[leftmargin=*]
    \item \textbf{Inference Engines.} Each engine manages one GPU.
    Engines are categorized into prefill engines (PEs) for prefill and decoding engines (DEs) for decode.
    \item \textbf{Traffic Manager (\autoref{sec:traffic-manager}).} Each engine contains a traffic manager to conduct (1) Host-Device memory copies (H2D \& D2H), (2) KV-Cache transfers between PEs and DEs, and (3) KV-Cache reads/writes from/to storage via the storage NIC. We adopt a CNIC-centric traffic management approach, detailed in \autoref{sec:traffic-manager}, to prevent KV-Cache traffic from affecting communications in model inference.
    \item \textbf{Request Scheduler (\autoref{sec:scheduling}).} A central scheduler that receives client requests and distributes them across engines. It is also responsible for dynamically distributing data traffic between two paths (\autoref{fig:dual-path-loading}).
\end{itemize}

\subsection{Dual-Path Loading}
\label{sec:dual-path-loading}

In addition to the conventional \emph{storage-to-prefill} path, \sysname introduces a novel \emph{storage-to-decode} path, allowing KV-Cache to be loaded first into a decode engine and then transferred to the prefill engine via high-bandwidth RDMA over the compute network. 
By dynamically distributing load across both paths, the system aggregates the storage NIC bandwidth of all engines --- including otherwise-idle decode-side NICs --- and eliminates the asymmetric bandwidth saturation that limits existing systems. This approach transforms the storage I/O from a single-bottleneck resource into a globally pooled and schedulable capacity. % The following subsections detail the data-flow design and prove its feasibility under typical deployment ratios.
The exact data flows of the dual-path are described below.

To implement dual-path loading, \sysname allocates a small amount of DRAM as buffers on each PE and DE, called \emph{PE buffer} and \emph{DE Buffer}.
% In \autoref{fig:pe_flow} and \autoref{fig:de_flow}, we ignore all newly computed KV-Cache, since its volume is bounded by the GPU's compute capacity and is therefore relatively small.

\parabf{Prefill PE read path.}
First, the KV-Caches of hit tokens are read from persistent storage into the PE buffer (as Label 1 and 2 shown in \autoref{fig:pe_flow}). 
Before the computation of an attention layer, those KV-Caches of that layer are transferred to PE HBM (3 and 4) to compute the KV-Cache of cache-miss prompt tokens.
Then, all KV-Caches of both hit and miss tokens are transferred to the DE buffer to form the complete prompt KV-Cache (5-7). This process (3-7) repeats $n_{layer}$ times.
During the prefill forward pass, transfers overlap with computation.

\parabf{Prefill DE read path.} The KV-Caches of hit tokens are first read into DE buffer (as Label 1 and 2 shown in \autoref{fig:de_flow}).
During PE prefill, KV-Cache for the corresponding layer is read from the DE buffer, also overlapping with computation (3-5). This process repeats $n_{layer}$ times.
After a layer's computation completes, only the KV-Caches of miss tokens are transferred to DE buffer and merged with the existing hit token KV-Cache.

\parabf{Decode Phase.} After receiving the complete prompt KV-Cache in DE buffer (including loaded KV-Cache via PE read path and the KV-Cache of newly appended tokens), the decode phase begins.
The DE first allocates HBM and performs host-to-device (H2D) transfers (Label 8 and 9 in \autoref{fig:pe_flow}; Label 6 and 7 in \autoref{fig:de_flow}), then releases CPU memory before starting decode.
The design of DE buffer imposes bandwidth pressure on DRAM and CNIC (an extra H2D), which could be avoided by directly bypassing it via GPU Direct RDMA.
However, since the generation length is typically short in this scenario, time-to-first-token (TTFT) accounts for a non-negligible portion of the total end-to-end request time.
Introducing DE buffer helps reduce GPU memory usage.
During decode, whenever a full block of tokens (e.g., 64 tokens) is accumulated, it is immediately persisted to disk.

\parabf{Different Block Layouts.}
We adopt two different block layouts: \emph{Full Block} and \emph{Layer Block}, which contain all layers and a single layer, respectively.
Detailed layout can be found in \autoref{subsec:block-layout}.
For all interactions with storage, we adopt Full Blocks.
In the PE read case, KV-Cache loading to PE HBM and transfer to DE Buffer occur in a layerwise streaming fashion, both using Layer Blocks.
Similarly, for the DE read path, transfers from the DE Buffer to the PE HBM use Layer Blocks.

\subsection{Bottleneck-Free Analysis}
\label{subsec:bottle-free-analysis}
We demonstrate that the system can fully saturate all storage NICs without introducing compute-NIC or DRAM bottlenecks, under most reasonable P/D ratios.
We assume a well-configured PCIe topology (each pair of GPU--NIC is under the same PCIe switch), load-balanced task scheduling, no congestion on the computation network, and that storage read bandwidth is fully utilized.

\parabf{Notation.} Let $P$ and $D$ denote the number of prefill and decode nodes, respectively. Each node has $g$ GPUs, each with one compute NIC of bandwidth $B$. The storage bandwidth per machine is $s \times B$ (shared by all engines on that machine); $M$ is the memory bandwidth per machine.

\parabf{Traffic per PE-DE pair.} We assume that the storage read bandwidth is fully utilized and that task scheduling is load-balanced. Under load balancing, storage NIC bandwidth is evenly shared. The traffic per pair for the PE read path (all steps in \autoref{fig:pe_flow}) is $T_p = Bs/(Dg^2)$; for the DE read path (\autoref{fig:de_flow}) it is $T_c = Bs/(Pg^2)$. Link traffic is the sum over all pairs using that link.

\parabf{PE CNIC Bandwidth Analysis.}
For PE CNIC, loopback traffic (i.e., H2D and D2H that does not traverse switches) exists, so the total traffic on the PCIe side is always greater than or equal to the switch-direction traffic, regardless of read or write operations.
Therefore, we only need to compute the pressure on the PCIe side.
Read operations include PE paths (3) and (5), with total traffic over all pairs:
\begin{align}
2\times T_p \times Dg = 2Bs/g \leq B
\end{align}
Since $s\leq g$ always holds in practice, the read direction is always bottleneck-free.
Write operations include PE path (4) and DE path (5), with total traffic:
\begin{align}
(T_p + T_c) \times Dg = Bs/g \times (1 + D/P) \leq B
\end{align}
Then, we obtain:
\begin{align}
P/D \geq \frac{s}{g-s}
\end{align}

\parabf{DE CNIC Bandwidth Analysis.}
For DE CNIC, read operations include PE path 8 and DE paths 3/6, with traffic:
\begin{align}
(T_p + T_c \times 2) \times Pg = s/g\times (P/D+2) \times B \leq B
\end{align}
Then, we obtain:
\begin{align}
P/D \leq \frac{g-2s}{s}
\end{align}

Write operations include PE paths 7/9 and DE path 7, with traffic:
\begin{align}
(2T_p + T_c) \times Pg \leq B
\end{align}
This implies:
\begin{align}
P/D \leq \frac{g-s}{2s}
\end{align}

\parabf{DRAM Pressure Analysis.}
DRAM is half-duplex, so we sum the read and write pressures.
For PE MEM, the pressure is $2sB$, which generally does not exceed memory bandwidth.
For DE MEM, following the similar analysis above, we can get the pressure is $(3 + 2P/D) Bs$.
Requiring the DE MEM pressure to be less than or equal to $M$, we obtain:
\begin{align}
P/D \leq \frac{M/Bs - 3}{2}
\end{align}

\parabf{Summary.} Combining all the above analyses, we have:
\begin{align}
\frac{s}{g-s} \leq P/D \leq \min\left\{\frac{g-2s}{s}, \frac{g-s}{2s}, \frac{M/Bs - 3}{2}\right\}.
\end{align}
For $(g=8, s=1)$ with $M \approx 500$ GB/s and $Bs \approx 50$ GB/s, the bottleneck-free range is $\frac{1}{7} \leq$ P/D $\leq \frac{7}{2}$, which covers most practical configurations.

\subsection{Practical Challenges}
\label{sec:system}

The dual-path architecture fundamentally reorients data movement: KV-Cache can be loaded either directly from storage into prefill engines or indirectly via decode engines, thereby aggregating storage bandwidth across all engines and breaking the prefill-side I/O bottleneck. However, realizing this high-level design in a practical system introduces three interrelated challenges. We briefly outline these challenges below and refer the reader to the corresponding sections for details.

\parabf{Fine-grained data transfer (\autoref{sec:traffic-manager}).} 
The layer-wise execution paradigm, while essential for overcoming HBM capacity limits, fragments the KV-Cache into numerous fine-grained blocks \cite{ISCA24:Splitwise}. Transferring this multitude of fine-grained data chunks between storage, host DRAM, and GPU HBM must incur minimal overhead and seamlessly overlap with computation to realize throughput gains. 

\parabf{Traffic isolation (\autoref{sec:traffic-manager}).} 
The complex data path in \sysname introduces additional KV-Cache transfer traffic on both the compute network and PCIe links. A primary concern is that this traffic may interfere with existing latency-sensitive collective communication operations essential for model execution --- such as AllToAll in expert parallel \cite{DeepEP} and ReduceScatter/AllGather in tensor/context parallel. Since these collective communications are critical to end-to-end inference latency, a key challenge lies in exploiting spare I/O bandwidth without degrading model inference performance.

\parabf{Dynamic load balancing (\autoref{sec:scheduling}). } 
As we are adopting two different paths for KV-cache loading, the system must promptly decide which path to use for each request. A naive policy could overload one path, recreating the original bottleneck. The traffic scheduler must balance multiple factors in real-time: storage NIC queue lengths, computational load on GPUs, and request workload characteristics.

\section{CNIC-Centric Traffic Manager}
\label{sec:traffic-manager}

Modern LLM inference systems employ a range of advanced data transfer technologies --- such as on-chip CUDA copy engine and GPUDirect Storage \cite{nvidia2026gds} --- to move data efficiently between storage, host memory, and GPU HBM. However, all these mechanisms can interfere with latency-sensitive collective communications (e.g., EP AllToAll) during model execution. This arises for two primary reasons: (1) such transfer technologies often operate over separate paths that do not share the same QoS controls as the compute network, and (2) existing GPUs do not support PCIe QoS \cite{7723586}, making it difficult to shield model inference communication from other traffic contending for PCIe bandwidth. Additionally, because the collective communications occur in rapid, sub-millisecond-level bursts, it is impractical to rely on a software-based traffic shaper to interleave lower-priority I/O operations between these high-priority traffic windows.

To address this, we propose a \textbf{CNIC--centric data transfer} approach which is widely adopted in our production deployment: all data traffic in or out of a GPU, including local H2D/D2H copy, must go through the GPU's paired CNIC with a GPUDirect RDMA \cite{nvidiaOverviewx2014} data path. By consolidating all traffic onto the compute network, we can leverage the native QoS capabilities of compute network to enforce strict traffic differentiation.

\subsection{Traffic Isolation}

For the InfiniBand-based network, we leverage virtual lanes (VLs) \cite{ibta-spec} to enforce isolation between different traffic classes. 
All model inference communication traffic is assigned to a dedicated high-priority VL, while all other traffic, including KV-Cache transfer, is mapped to a separate low-priority VL.
We configure the VL arbiters of all network switches and NICs with a weighted round-robin policy that reserves approximately 99\% of total bandwidth to high-priority VL.
The remaining bandwidth is allocated to the low-priority VL to prevent starvation.
This configuration ensures that model execution traffic is virtually unaffected by KV-cache transfers, while still allowing KV-cache traffic to opportunistically utilize otherwise idle bandwidth in the compute network.
Detailed configurations are described in \autoref{sec:traffic-isolation-config}.

Although our experiments are conducted on an InfiniBand-based network, the same design principles naturally extend to other interconnect technologies.  
\sysname can be conducted on RDMA over Converged Ethernet (RoCE) by leveraging Traffic Class (TC) and Differentiated Services Code Point (DSCP) markings \cite{sigcomm16:rocev2,Carpenter2002DifferentiatedSI} in conjunction with hardware packet queues.  
Emerging technologies such as UnifiedBus \cite{unifiedbus} and Ultra Ethernet \cite{uec-spec} are likewise converging on QoS mechanisms for heterogeneous traffic, which can directly support the requirements of \sysname.

\subsection{CNIC-Assisted KV-Cache Copy}

Existing GPU data transfer technologies include GPUDirect Storage \cite{nvidia2026gds}, which loads KV-Cache from storage backend to GPU HBM, and CUDA copy engine, which directly copies host DRAM to GPU via PCIe. However, these methods fail to isolate KV-Cache traffic from high-priority latency-sensitive collective communications in model execution, severely degrading inference performance.

To solve the limitations of existing approaches, we adopt a CNIC-assisted H2D/D2H data path. For KV-Cache loading, we first read the KV-Cache into host DRAM from the storage backend. Then, we submit an RDMA Write request to the GPU's paired CNIC to perform local H2D copy. Storing newly generated KV-Cache follows a symmetric process: it is first transferred to host DRAM via CNIC, then persisted to the storage backend over the storage network. This design establishes the CNIC as the central QoS scheduler for all GPU PCIe traffic, allowing its VL arbiter to prioritize the inference communication traffic and perform KV-Cache transfer using spare PCIe bandwidth. 

Although this approach may appear to take a detour compared to GPUDirect Storage (which directly reads KV-Cache to GPU HBM) and CUDA copy engine (which directly copies host memory to GPU HBM), to our best knowledge, this is currently the only practical method to ensure that KV-Cache load/store does not degrade the performance of critical model--execution communication.

We also observe that CNIC-assisted H2D and D2H outperform the CUDA copy engine when handling a large number of small data chunks. Our measurements show that submitting a single copy operation via \texttt{cudaMemcpyAsync} incurs a latency overhead of approximately {5}-{7}$\mu s$. We failed to further break down this overhead due to the closed-source nature of CUDA driver. In contrast, submitting one RDMA Write work request involves only a few mmio writes to NIC registers in user space and takes only around {1}$\mu s$. Furthermore, the RDMA work submission overhead can be significantly amortized by leveraging \emph{doorbell batching} \cite{kalia2016design}.

\section{Adaptive Request Scheduler}
\label{sec:scheduling}

Although our theoretical analysis shows promising results, imbalanced load reduces hardware utilization, and in this scenario, we need to consider two dimensions of balance simultaneously: (1) NIC traffic, and (2) the utilization balance of GPUs.
We divide scheduling into two levels: inter-engine scheduling, which assigns requests to a (PE, DE) pair and selects the read path (PE or DE) for each request; and intra-engine scheduling, which determines which requests are included in each forward batch for computation.

\subsection{Inter-Engine Scheduling}

We organize engines into groups to reduce the scheduler pressure.
Only the engine rank $0$, called \emph{Leader Engine}, interacts with the scheduler.
All engines of a group are all PEs or all DEs.
All engines on one node are guaranteed to be in the same group.
All engines in one group proactively fetch tasks together regularly.
When fetching new requests, each engine $e$ reports (1) $seq_e$, the number of requests assigned to it that have not yet completed; (2) the total token count $tok_e$ over those $seq_e$ requests; and (3) the disk reading queue length $read\_q_{n(e)}$ of the node $n(e)$ that engine $e$ belongs to.
GPU load, disk read load, and network load are all strongly correlated with token count. We therefore use token count as a proxy and aim to balance it across engines.

\begin{figure}[t!]
  \centering
  % \hspace*{-0.85cm}
  \includegraphics[width=\linewidth]{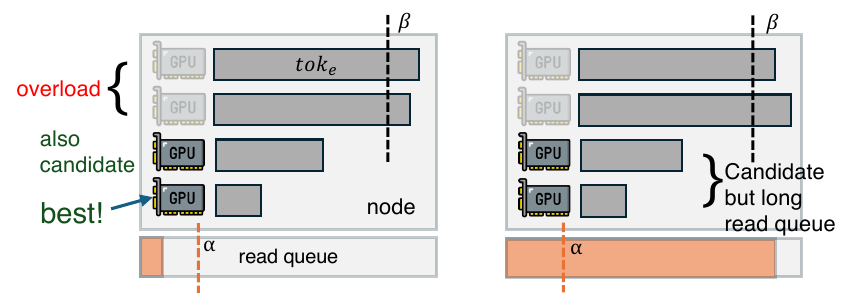}
  \vspace{-0.2in}
  \caption{\normalfont An illustration of Inter-Engine PE Scheduling. All eight GPUs are in the same PE engine group and the scheduler will choose the best.}
  \vspace{-0.2in}
  \label{fig:intersched}
\end{figure}

\parabf{PE Scheduling.} All requests arriving at the scheduler enter a waiting queue and are scheduled in a FIFO order. The scheduling algorithm is invoked when a PE group initiates a fetch request.
An illustration of the inter-engine scheduling process is shown in \autoref{fig:intersched}.
We define two constants, short reading queue threshold $\alpha$, and unfinished token upper limit $\beta$, measured in tokens.
All engines are split into three categories:
(1) overloaded engines where $tok_e > \beta$;
(2) engines on nodes with short disk reading queues where $read\_q_{n(e)} \leq \alpha$ and $tok_e \leq \beta$; and
(3) engines on nodes with longer disk reading queues where $read\_q_{n(e)} > \alpha$ and $tok_e \leq \beta$.
We do not assign new requests to overloaded engines.
Second-category engines are prioritized over third-category engines because they reside on nodes with shorter disk reading queues, and lack of subsequent requests would easily lead to storage NIC underutilization.

We assign the current request to the PE with minimum $tok_e$ in the second category if non-empty, otherwise in the third category if non-empty.
After assignment, we update the selected PE's $tok_e$, then proceed to the next request in the waiting queue.
If both categories are empty, we terminate this fetch request and return the already-assigned requests to the Leader Engine.

\begin{algorithm}[t]
\caption{\normalfont Inter-PE Scheduling Algorithm}
\label{alg:pe-scheduling}
\SetAlgoLined
\KwData{Waiting queue $Q$, PE group $G_{PE}$, load metrics $(read\_q_{n(e)}, tok_e)$ reported by each engine $e$, where $n(e)$ denotes the node that engine $e$ belongs to, constants $\alpha$ and $\beta$}
\KwResult{Assigned requests to PEs}
\BlankLine
Each engine $e$ reports $(read\_q_{n(e)}, tok_e)$\;
Classify all PEs into three categories:\;
\Indp $C_1 \leftarrow \{e \in G_{PE} : tok_e > \beta\}$\;
$C_2 \leftarrow \{e \in G_{PE} : read\_q_{n(e)} \leq \alpha \wedge tok_e \leq \beta\}$\;
$C_3 \leftarrow \{e \in G_{PE} : read\_q_{n(e)} > \alpha \wedge tok_e \leq \beta\}$\;
\Indm
\While{$Q$ is not empty}{
    $r \leftarrow$ head of $Q$\;
    \eIf{$C_2 \neq \emptyset$}{
        $pe^* \leftarrow \arg\min_{e \in C_2} tok_e$\;
    }{
        \eIf{$C_3 \neq \emptyset$}{
            $pe^* \leftarrow \arg\min_{e \in C_3} tok_e$\;
        }{
            Terminate this fetch request\;
            Return assigned requests to Leader Engine\;
            \textbf{break}\;
        }
    }
    Assign request $r$ to PE $pe^*$\;
    Update $tok_{pe^*} \leftarrow tok_{pe^*} + \text{tokens}(r)$\;
    Remove $r$ from $Q$\;
}
\end{algorithm}

\parabf{DE Scheduling Phase 1: across groups.}
DE scheduling is two-level and does not preserve global FIFO. There is a global waiting queue and a private queue per DE engine group. Incoming requests first enter the global queue. When a DE group fetches, \emph{group-level} scheduling drains the global queue and assigns each request to the group whose total $tok_e$ (sum over its engines) is minimum; this balances token count across groups and thus NIC and GPU load. 

\parabf{DE Scheduling Phase 2: within a group.}
Then we calculate the sum of remaining HBM for all DEs in the group and traverse from the head of the private queue to calculate how many requests can be scheduled assuming no HBM fragmentation. These requests form the set $R$.
It is an upper bound that can be scheduled. 
Then, we calculate a high token threshold $Z = 1.05 \times (\sum_{r\in R} {len_r} + \sum_{e\in E} {tok_e}) / |E|$.

Next, we try to pop the head of private queue and schedule it to a DE. 
Among DEs with sufficient remaining HBM for the request, we partition into (1) high-token DEs, where $tok_e + \text{len}(r) > Z$, and (2) the rest. We prefer category (2) to keep token count balanced; category (1) DEs already have higher GPU and NIC pressure. Within category (2) we choose the DE with minimum $seq_e$ to balance request count; if category (2) is empty we choose the DE with minimum $tok_e$ in category (1) to reduce HBM exhaustion and preemption risk. If no DE has sufficient HBM, the fetch ends and already-assigned requests are returned.

\parabf{KV-Cache Read Task Scheduling.} 
After selecting PE and DE for a request, we choose to read on the side with the shorter reading queue. It is probably better to split the request into two parts and read them from both sides, and we leave it as future work.

\subsection{Intra-Engine Scheduling}

\begin{figure}[t!]
  \centering
  \includegraphics[width=\linewidth]{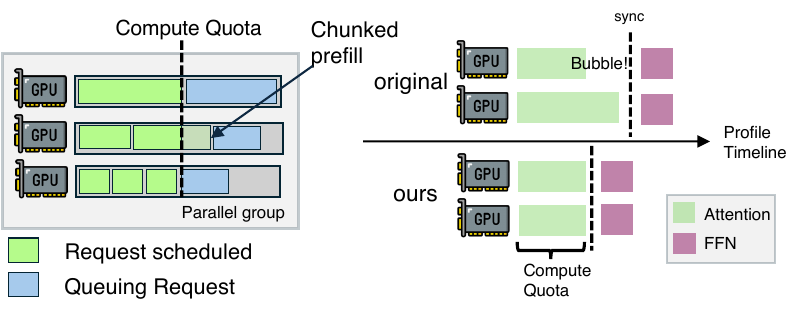}
  \caption{\normalfont Intra-Engine Schedule. Left: compute-quota-based batch selection. Right: GPU timeline before and after applying compute quota.}
  \label{fig:intrasched}
\end{figure}

Only PEs require intra-engine scheduling, as DEs always place all requests into the forward batch.
An illustration of the intra-engine scheduling process is shown in \autoref{fig:intrasched}.
Data parallelism is widely adopted for attention layers, especially for MLA models.
Under such a parallel configuration, each GPU serves a different set of requests. 
It may lead to workload imbalance among all GPUs that must synchronize after the attention stage and enter the FFN stage together, causing GPU bubbles waiting for other peers.
Therefore, we need to make sure they have similar attention layer execution times to minimize the waiting bubbles.

\parabf{Layer Time Estimation.} We use FIFO packing to decide how many requests to include in a forward batch.
Each request in a forward batch is described by a pair $(cached, bsz)$, where $cached$ is the number of tokens with KV-Cache already available (from storage hits or previous forward passes), and $bsz$ is the number of tokens requiring KV-Cache computation in this forward batch.
From these pairs, we compute the total theoretical computation for the attention layer and estimate its execution time.
The relationship between theoretical computation and wall-clock time depends on hardware and parallel configuration, and can be fitted in advance through profiling as previous works \cite{SOSP25:PrefillOnly} and \cite{OSDI24:SarathiServe}.

\parabf{Algorithm.} We keep adding requests in FIFO order as long as the predicted attention layer execution time does not exceed a predefined upper bound, called the \emph{compute quota}.
If adding a request would exceed this bound, we perform binary search on $bsz$ to find a smaller $bsz'$ to fit in the remaining compute quota and perform chunked prefill for that request. 

% \FloatBarrier

\section{Evaluation}
\label{sec:evaluation}

\begin{figure*}[htbp]
  \centering
  \includegraphics[width=0.98\linewidth]{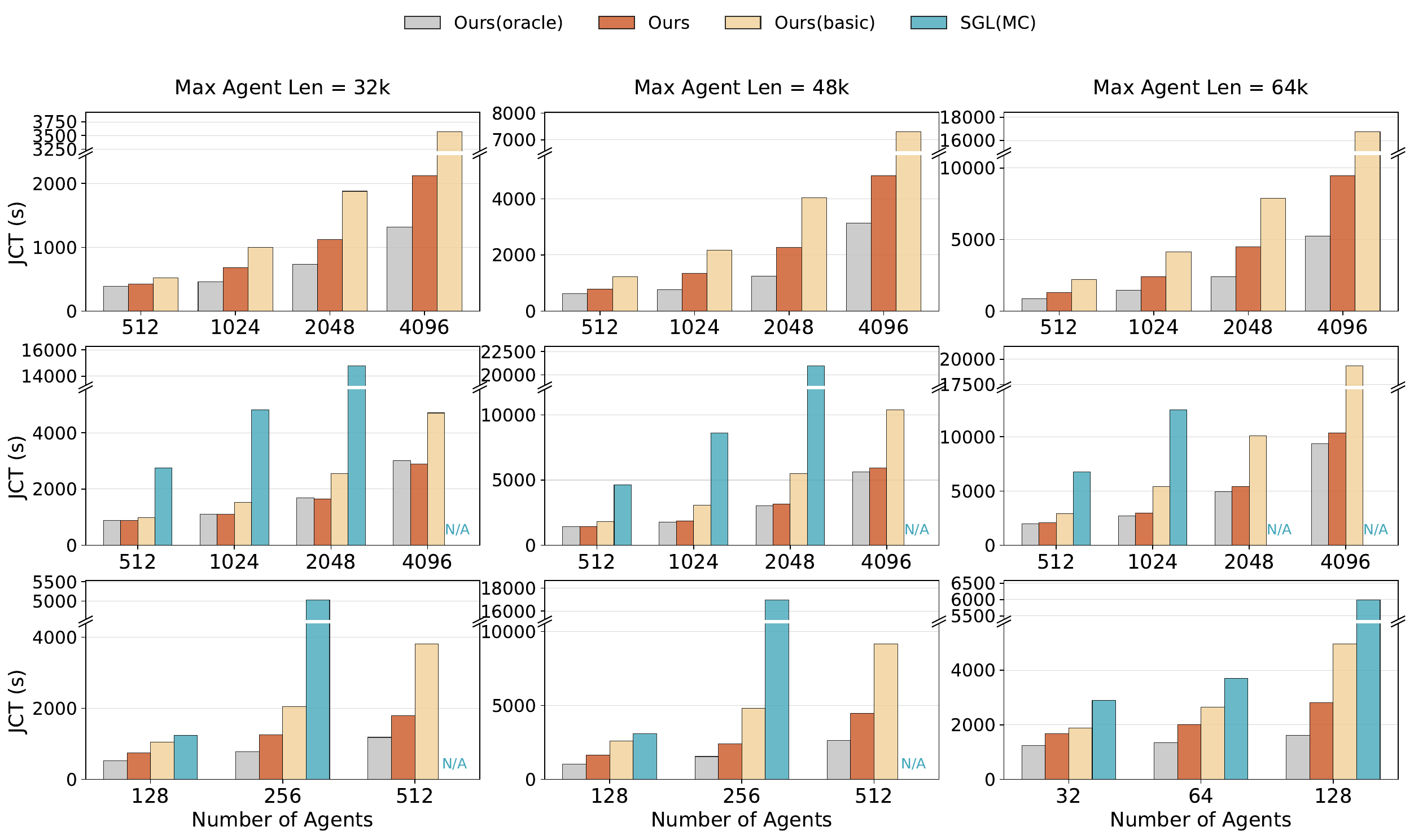}
  \caption{\normalfont Offline inference performance under varying numbers of agents and maximum agent context lengths. Top: DS 27B. Middle: DS 660B. Bottom: Qwen 32B. N/A for running into an error before finishing.}
  \label{fig:rollout}
\end{figure*}

\subsection{Implementation}
\label{subsec:impl}
We implement \sysname based on our in-house inference framework.
For CUDA kernels, our in-house framework adopt the combination of FlashMLA \cite{FlashMLA}, DeepGEMM \cite{DeepGEMM}, and DeepEP \cite{DeepEP}, which aligns with the current mainstream open-source framework \cite{NIPS24:SGLang, SOSP23:PagedAttention}.
The \sysname implementation involves approximately 5K lines of modifications on top of it.
We adopt 3FS \cite{3FS} as distributed storage and use an \verb|io_uring|-like interface for kernel bypass.

\subsection{Experimental Setup}
\label{subsec:experimental-setup}

\paraf{Testbed.}
We conduct our experiments on a cluster of GPU servers with InfiniBand interconnection. Each server has 8 NVIDIA Hopper GPUs and dual processors. Additionally, each node is provisioned with eight 400Gbps RDMA NICs connected to InfiniBand network and one additional storage NIC connected to 3FS. The computation and storage networks are physically isolated. Our cluster-wide 3FS has no internal DRAM cache and can saturate the 400Gbps bandwidth of the storage NIC.

\parabf{Models.}
We evaluate on three models: (1) DeepSeek V3.2 \cite{DeepSeekV32} 660B, an MoE model with DeepSeek Sparse Attention, denoted as \emph{DS 660B}, (2) a 27B downscaled version of DS 660B, denoted as \emph{DS 27B}, and (3) Qwen2.5-32B \cite{qwen2025qwen25technicalreport}, a dense model with GQA, denoted as \emph{Qwen 32B}.
DS 660B and Qwen 32B correspond to the publicly released checkpoint on HuggingFace.
DS 27B is our internal experimental model with a similar architecture to DS 660B.
Detailed specifications are provided in \autoref{sec:27b-model-specs}.

\parabf{Datasets.}
We collected three agent trace datasets from our production agentic RL training workloads with varying maximum context lengths (MaxLen). Each dataset contains 500 trajectories. The average interaction turns (Turns), average appended and generated tokens per turn (Append and Gen), average number of total tokens (Total), and average number of context tokens (Context) are summarized in \autoref{tab:agent-dataset-statistics}. 

\begin{table}[t!]
\centering
\caption{\normalfont Statistics of agent trace datasets.}
\begin{tabular}{cccccc}
\toprule
\textbf{MaxLen} & \textbf{Turns} & \textbf{Append} & \textbf{Gen} & \textbf{Total} & \textbf{Context} \\
\midrule
32K  &  60 & 608  & 148 & 28639 & 17183 \\
48K  & 106 & 474  & 172 & 42607 & 25120 \\
64K  & 157 & 429  & 176 & 55958 & 32721 \\
\bottomrule
\end{tabular}
\label{tab:agent-dataset-statistics}
\end{table}

\parabf{Baselines.}
We compare \sysname, denoted as Ours, against the following baselines:

\begin{itemize}[leftmargin=*]
\item \textbf{SGL(MC)}: SGLang \cite{NIPS24:SGLang} (commit 19089aa) with HiCache \cite{Hicache}, Mooncake \cite{FAST25:Mooncake} Store enabled and 3FS as the storage backend, and Mooncake Transfer Engine for prefill-decode disaggregation.
We did not run SGL(MC) for DS 27B because SGLang lacks support for this downscaled version.
\item \textbf{Basic}: Our unmodified internal inference framework (detailed in \autoref{subsec:impl}).
Comparing \sysname and SGL(MC) is unfair due to implementation differences.
Therefore, we only report performance improvements from Basic to Ours.
\item \textbf{Oracle}: Based on \sysname, we bypass all disk reads, D2H \& H2D transfers, and inter-PD KV-Cache transfers.
This configuration represents the theoretical performance upper bound assuming zero I/O overhead.
\end{itemize}

\parabf{P/D Ratio and Parallelism.} 
We default to 2P4D for DS 660B, 1P2D for Qwen 32B, and 1P1D for DS 27B (where 1P1D means one node for each side).
For DS models, we use EP and DP.
For Qwen 32B, we use DP only in \sysname, while SGL(MC) uses TP=8 since DP attention is not supported for this model in SGLang.
Detailed configuration is provided in \autoref{sec:experimental-configurations}. 

\parabf{Metrics.} For offline inference scenarios, we measure job completion time (JCT) for the entire task.
For online serving scenarios, we measure TTFT, TTST (Time to the second token), and TPOT.

\subsection{Offline Batch Inference}
\label{subsec:offline-batch-infer}

This section evaluates throughput performance in offline batch inference, which is the case of the rollout phase in RL training.
In this scenario, $n$ agents start to rollout simultaneously, and we measure the JCT when all requests have finished.

\begin{figure*}[htbp]
  \centering
  \includegraphics[width=\linewidth]{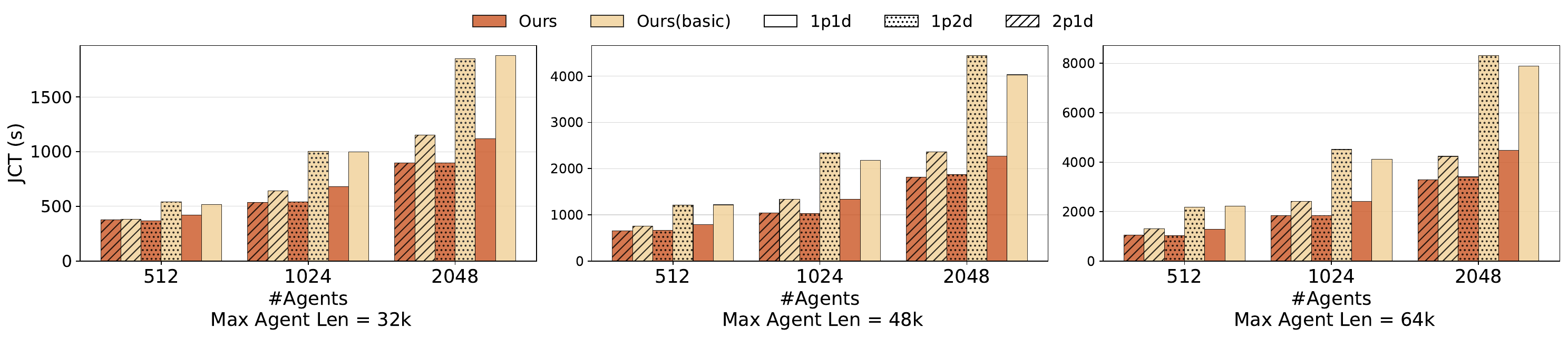}\\
  \caption{\normalfont Impact of prefill-decode ratio on offline inference performance (DS 27B).}
  \label{fig:rollout_diff_pd}
\end{figure*}

\parabf{Varying Agents Batch Size \& Max Agent Length (MAL).}
\sysname benefits more from larger batch sizes and longer MALs.
\autoref{fig:rollout} reports JCT under different batch sizes and MALs. 
SGL(MC) encountered errors in our setup and failed to complete some large configurations (marked as N/A).
On DS 660B, \sysname achieves up to $1.87\times$ over Basic, and demonstrates performance with Oracle, indicating that KV-cache I/O is largely eliminated.
On DS 27B, \sysname improves over Basic by up to $1.78\times$ but remains $1.09$--$1.85\times$ slower than Oracle due to limited storage bandwidth in 1P1D (\autoref{fig:rollout_diff_pd}).
For Qwen 32B, it shows similar trends as DS 27B.

\begin{figure}[t]
  \centering
  \includegraphics[width=\linewidth]{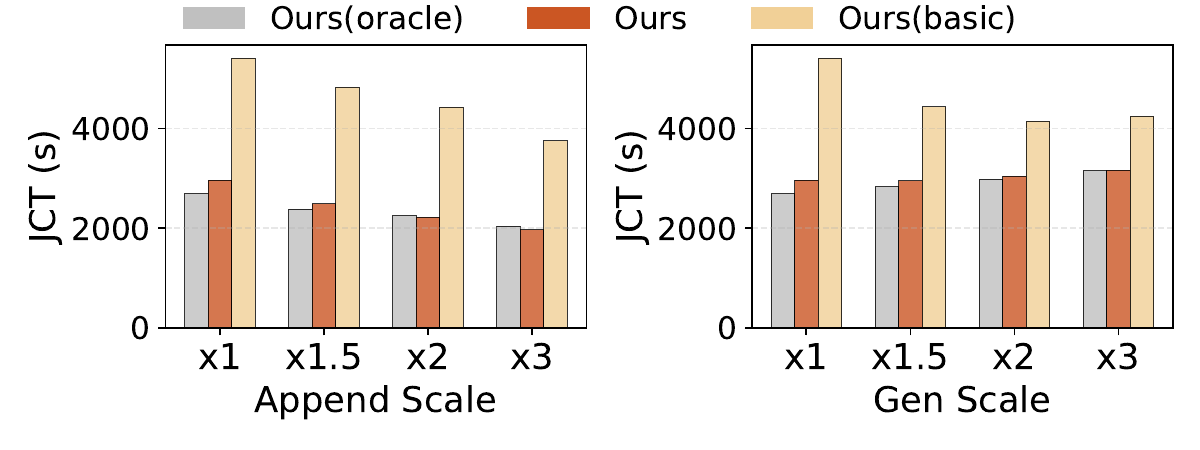}
  \caption{\normalfont Left: varying append lengths (DS 660B, 64K context, 1024 agents). Right: varying generation lengths (DS 660B, 64K, 1024 agents)}
  \label{fig:rollout_diff_append}
  \end{figure}

\begin{figure*}[htbp]
  \centering
  \includegraphics[width=0.88\linewidth]{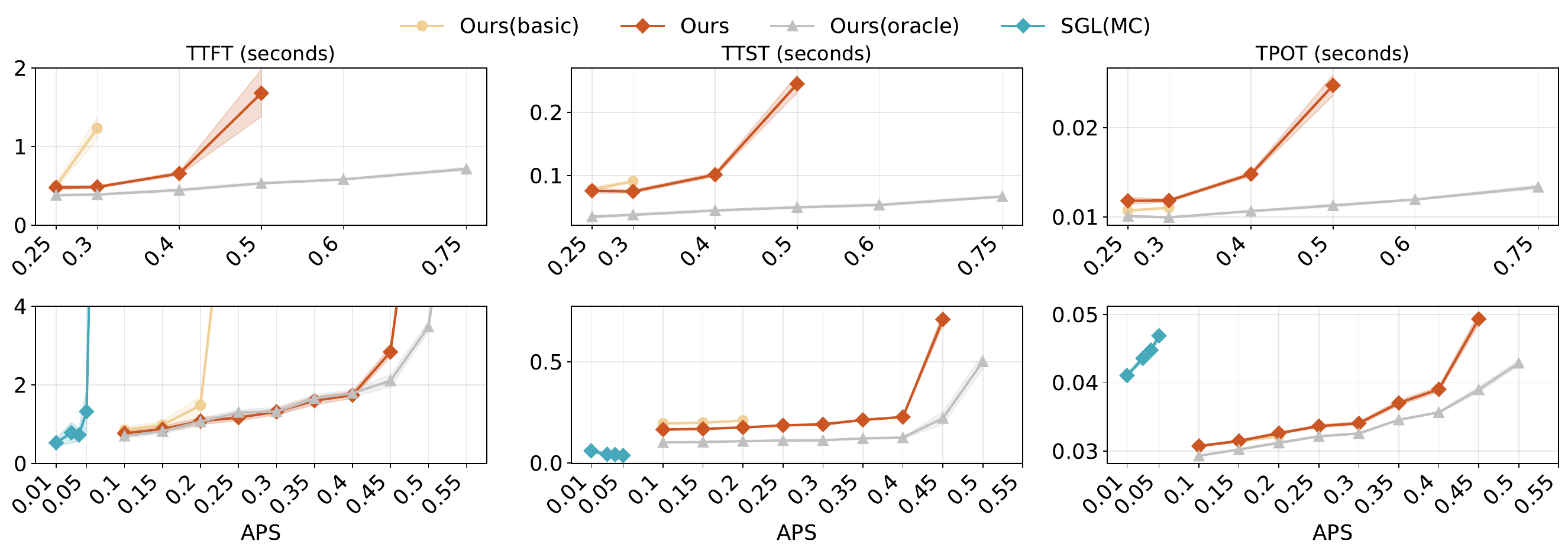}
  \vspace{-2mm}
  \caption{\normalfont TTFT, TTST, and TPOT as functions of agent arrival rate (APS). Shadow means the fluctuation in the last 150s before experiments finish. Top: DS 27B, Bottom: DS 660B.}
  \label{fig:serving}
\end{figure*}

\parabf{Varying Append Length \& Generation Length.}
\sysname has more advantages when append and generation tokens are short.
Longer append lengths imply greater GPU compute pressure, and longer generation lengths lower KV-Cache loading pressure due to larger prefill gap time.
To investigate the impact of this factor, we scale each round's append length by a constant factor, and then truncate the whole trajectory at given MAL.
The same holds for generation length.
As shown in \autoref{fig:rollout_diff_append}, with append length increases, Basic performance gradually approaches \sysname and Oracle, while \sysname and Oracle performance changes only slightly, indicating that the bottleneck consistently lies in GPU compute pressure.
Compared to Basic, \sysname achieves $1.82-1.99\times$ speedup at different append scales.
The trend for generation length scaling is similar.

\parabf{Varying Prefill-Decode Ratio.}
Across all ratios, \sysname demonstrates substantial performance gains compared to Basic.
We conduct rollout experiments on DS 27B with 1P1D, 2P1D, and 1P2D prefill-decode ratios to characterize the impact of resource allocation between prefill and decode stages on overall system performance.
As shown in \autoref{fig:rollout_diff_pd}, \sysname achieves an average speedup of $1.64\times$ across all configurations (up to $2.46\times$).
Basic 1P1D and Basic 1P2D perform comparably; 
so do \sysname 1P1D and Basic 2P1D, 
as well as \sysname 2P1D and \sysname 1P2D.
This occurs because each pair of systems has equivalent available storage bandwidth (Basic can only use prefill node storage bandwidth, while \sysname can utilize all nodes), which confirms that storage bandwidth is the dominant bottleneck in agentic scenarios.

\subsection{Online Serving}
\label{subsec:online-serving}
\parabf{Methodology.} We evaluate system latency characteristics under varying agent arrival rates per second (APS).
Agents arrive according to a Poisson process at a specified rate, with each agent commencing replay from round zero to its last round upon arrival.
For our experiments, the SLO is set as TTFT $\leq$ 4 seconds and TPOT $\leq$ 50ms.
In the TPOT and TTST figures, data points exceeding the SLO threshold are omitted.
Experiment termination is triggered when either: (1) TTFT exceeds 4 seconds, or (2) the system reaches steady state, defined as TTFT variation within a 150-second sliding window remaining below 5\% compared to that 30 minutes prior.

As shown in \autoref{fig:serving}, \sysname achieves higher APS capacity than Basic (1.67$\times$ for DS 27B, 2.25$\times$ for DS 660B).
\sysname's TTST is comparable to Basic, while TPOT shows that \sysname does not introduce additional decoding overhead compared to Basic.
SGL(MC) exhibits anomalously low TTST, likely due to implementation issues where the first two tokens arrive at the 
client almost simultaneously.
For DS 27B, all metrics exhibit trends similar to DS 660B.
However, both Basic and \sysname show significantly higher TPOT than Oracle, suggesting the overhead of basic P-D transferring is considerable in small model cases. We leave it as future work.

\begin{figure}[hbtp]
  \centering
  \includegraphics[width=0.9\linewidth]{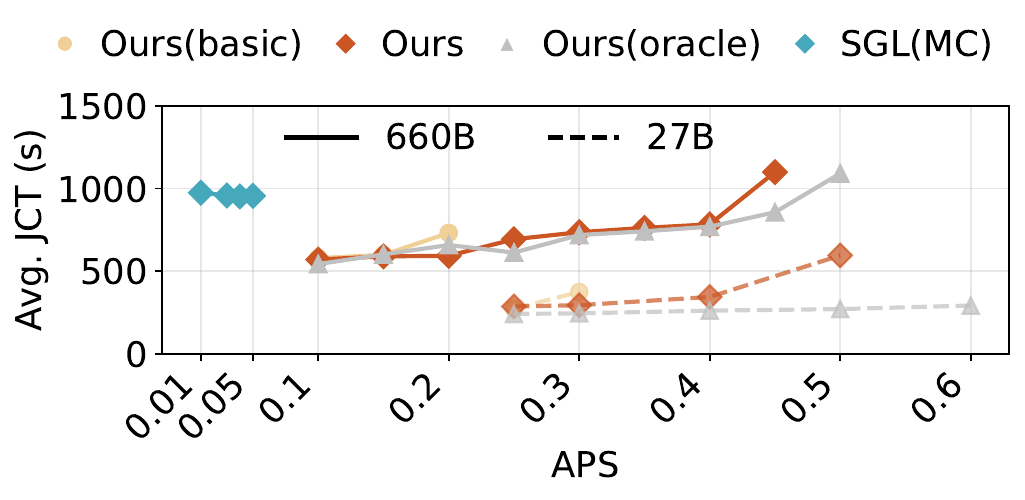}
  \vspace{-2mm}
  \caption{\normalfont Average completion time of all trajectories versus arrival rate for online serving.}
  \label{fig:serving-jct}
\end{figure}
Average JCT for both models are presented in \autoref{fig:serving-jct}.
A detailed analysis of working set implications is discussed in \autoref{sec:discussion}.
As shown in \autoref{fig:ablation-and-ttft-breakdown} (left), \sysname maintains stable TTFT components across different APS, while Basic's queuing time grows dramatically due to insufficient storage bandwidth.

\subsection{Ablation Study}
\label{subsec:ablation}

\begin{figure}[t]
  \centering
  \includegraphics[width=\linewidth]{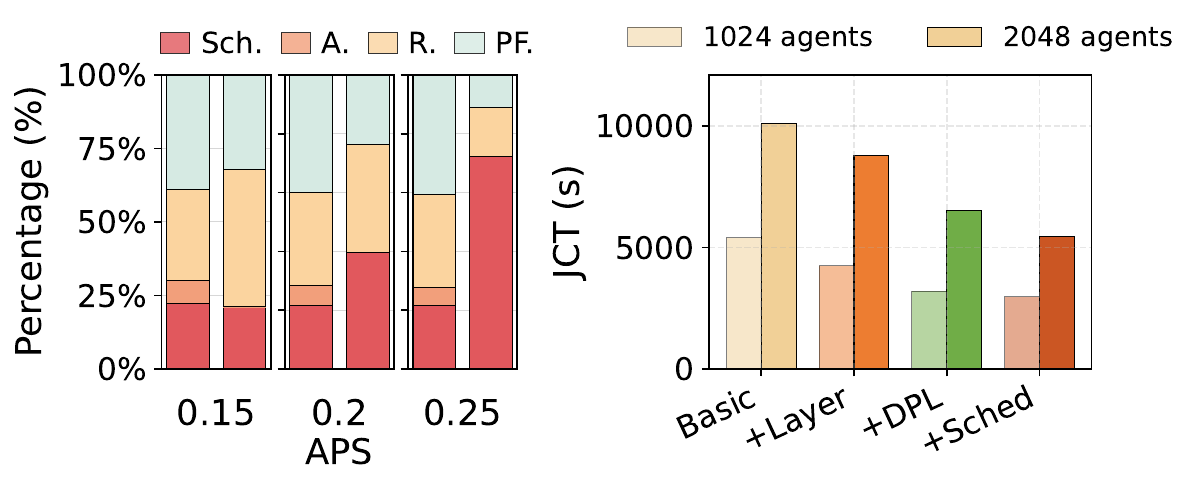}
  \caption{\normalfont Left (\autoref{subsec:online-serving}): TTFT breakdown for online serving (DS 660B) across APSs, Sch. for scheduling, A. for allocating, R. for reading KV-cache, PF. for prefill. In each pair of pillars, the first is for \sysname and the second is for Basic. Right (\autoref{subsec:ablation}): Offline inference ablation results (DS 660B, 64K context length). Layer, DPL, Sched stands for Layerwise prefill, Dual-Path Loading, and scheduling, respectively.}
  \label{fig:ablation-and-ttft-breakdown}
\end{figure}

We conduct an ablation study to quantify the contribution of each technical component in \sysname.
Experiments are performed under the offline inference setting with 64K MAL and agent batch size 1024 and 2048.
The differences between Basic and Ours are grouped into three techniques: layerwise prefill, dual-path loading, and scheduling algorithm. 
We add the techniques gradually to demonstrate individual contribution.
As shown in \autoref{fig:ablation-and-ttft-breakdown}, compared to Basic, adding layerwise prefill reduces JCT by $17.21\%$ on average, alleviating PE HBM bottlenecks and hiding transfer overhead.
Adding Dual-path loading on top of layerwise prefill delivers the primary performance gains, reducing JCT by $38.19\%$ on average compared to Basic, as it enables requests to read KV-Cache from either PE or DE, fully utilizing distributed storage bandwidth.
Finally, employing our scheduling algorithm on top of dual-path loading to decide KV-Cache loading paths achieves the best performance, reducing JCT by $45.62\%$ compared to Basic, demonstrating the effectiveness of load-balanced scheduling across storage NICs.

\begin{figure}[t]
  \centering
  \includegraphics[width=0.9\linewidth]{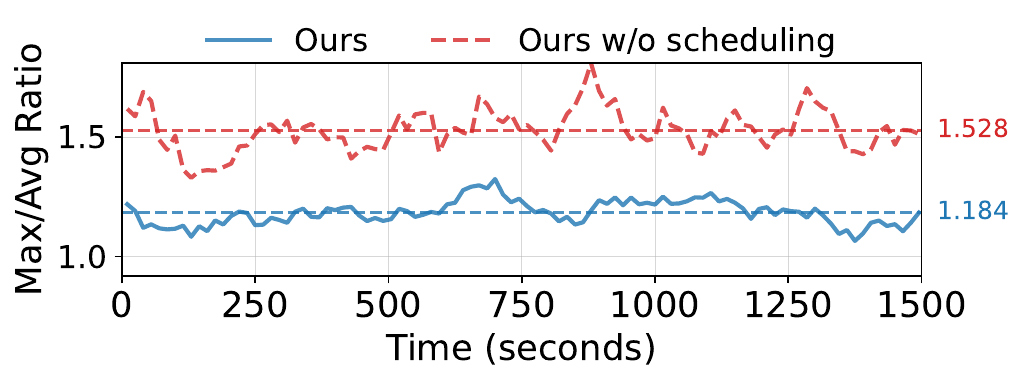}
  \vspace{-3mm}
  \caption{\normalfont Load balance of storage NICs traffic}
  %  with 64k MAL, 1024 agents batch size.
  \label{fig:read-lb}
\end{figure}

\begin{figure}[t]
  \centering
  \includegraphics[width=\linewidth]{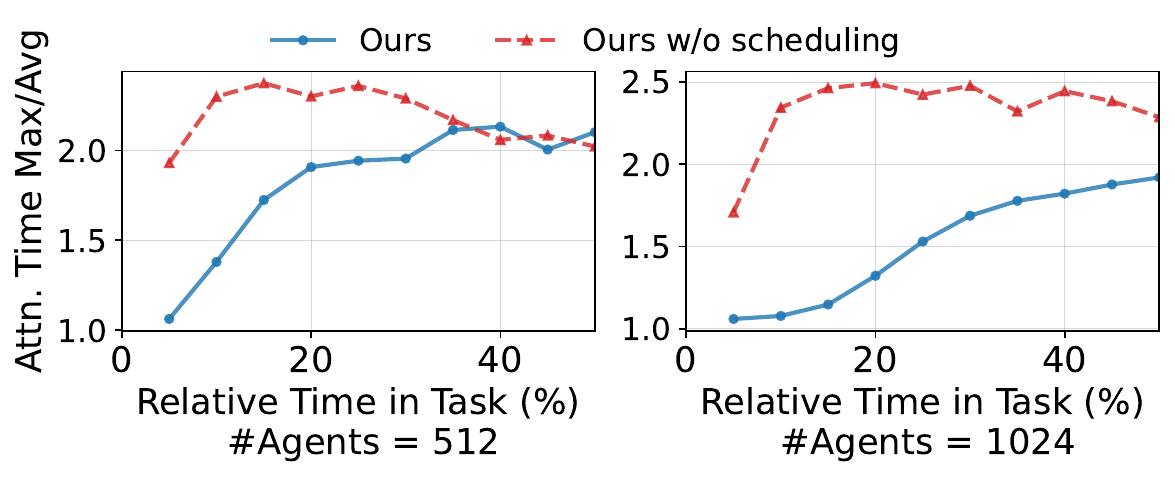}
  \caption{\normalfont Load balance of attention execution time.}
  \label{fig:attn-lb}
\end{figure}

\parabf{Load Balance.}
\sysname's scheduling algorithm improves load balance for both storage NICs and attention layer execution times.
For storage NICs, our scheduling algorithm improves load balance from 1.53 to 1.18 compared to round robin scheduling (\autoref{fig:read-lb}).
For attention layers, \sysname maintains the Max/Avg ratio as low as 1.06 during the first 5\% of the task, reducing GPU idle bubbles (\autoref{fig:attn-lb}).
The storage NIC metric is the ratio of maximum to average traffic across all storage NICs on three machines within a small time window, where 1.0 represents perfect balance.
The attention layer metric is calculated among all GPUs in an expert parallel group for each forward.
As the task progresses, both ratios become meaningless due to underloaded system. 
Therefore, we do not show the tail phase of the workload.

\begin{table}[t]
  \centering
  \caption{\normalfont Large-scale experiment results.}
  \label{tab:largescale}
  \setlength{\tabcolsep}{3pt}
  \small
  \begin{tabular}{cccccc}
  \toprule
  & \textbf{Setting} & \textbf{JCT} & \textbf{TTFT} & \textbf{TTST} & \textbf{TPOT} \\
  \midrule
  \multirow{2}{*}{\rotatebox{90}{\scriptsize Offline}} 
  & 2P4D, 2K agents & 3,167s & -- & -- & -- \\
  & 48P96D, 48K agents & 3,201s & -- & -- & -- \\
  \midrule
  \multirow{2}{*}{\rotatebox{90}{\scriptsize Online}} 
  & 2P4D, 0.4 APS & -- & 1.739s & 0.228s & 0.039s \\
  & 44P88D, 8.8 APS & -- & 1.847s & 0.194s & 0.036s \\
  \bottomrule
  \end{tabular}
  \end{table}

\begin{figure}[t]
  \centering
  \includegraphics[width=\linewidth]{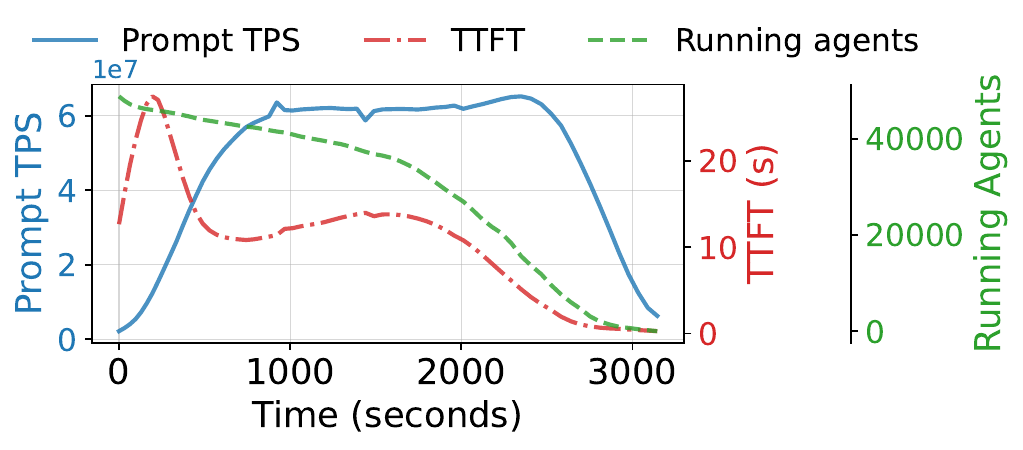}
  \caption{\normalfont 48P96D offline inference metrics. 1e7 is the scaling factor of Prompt TPS.}
  \label{fig:largescale-rollout}
\end{figure}

\subsection{Large-Scale Scalability}

We conduct both offline and online experiments using up to 1,152 GPUs to demonstrate large-scale scalability (\autoref{tab:largescale}). 
For offline inference, scaling from 2P4D (2K agents) to 48P96D (48K agents) achieves near-linear speedup with comparable JCT (3,167s vs.\ 3,201s). 
For online serving, the 44P88D configuration achieves 22$\times$ throughput (8.8 vs.\ 0.4 APS) while maintaining similar latency. 
Across all experiments, scheduler CPU usage remains below 10 cores, confirming it is not a bottleneck. Some detailed metrics over offline inference process are shown in \autoref{fig:largescale-rollout}.

Due to the lack of fine-tuned parallelism settings and P/D ratios (which requires a substantial experimentation budget), the large-scale experiments do not demonstrate additional JCT or serving capacity gains compared to multiple small-scale units with equivalent cost.
However, large-scale setting remains important for the following reasons. First, it reduces fragmentation and provides greater flexibility for fine-tuning parallelism and P/D ratios. Second, large-scale deployment offers more scheduling opportunities to mitigate queuing latency under unpredictable bursty online requests. These observations suggest several directions for future work (\autoref{subsec:futurework}).

\section{Discussion}
\label{sec:discussion}

\subsection{Potential Future Work}
\label{subsec:futurework}

The workload of offline inference is highly dynamic. For example, in our agentic RL tasks, the workload depends heavily on the researchers' algorithm design, and the prefill stage typically experiences significantly higher pressure in the first half of execution than in the second half. Meanwhile, profiling these tentative experiments is costly, as some experiments are only run a limited number of times. Therefore, more adaptive and flexible approaches for parallelism and P/D ratio configuration are needed, such as simulators or online adjustment mechanisms. Second, the scheduling algorithm still has room for improvement, as we expect to achieve lower TTFT percentiles under large-scale deployment.

\subsection{Working Set Analysis}

As shown in \autoref{fig:serving-jct}, given an arrival rate $\lambda$ (i.e., new trajectories per second) and mean JCT $\bar{T}$, the working set of the KV-Cache can thus be approximated as $\lambda \bar{T} \times total\_len_{avg}/2$.
In our setting of DS 660B serving, this value of \sysname ranges from 69 GB at APS 0.1 to 681 GB at APS 0.45.

In realistic setting, the working set would be larger since our evaluation assumes zero inter-arrival time and zero tool call latency. 
If JCT increases by $r$ times due to these gaps, the system's APS capacity increases by $r$ times (gaps do not stress LLM inference), causing the working set to expand by $r^2$ times.
This would exceed available memory and reduce the distributed memory pool's hit rate.
Such experiments require $r$ times more machine hours and $r^2$ times more storage (cost scaling as $r^3$), which we cannot afford given limited resources.

\section{Related Work}

\paraf{Distributed Memory Cache Pools.}
Mooncake~\cite{FAST25:Mooncake} builds a distributed DRAM pool for KV-Cache.
TokenLake~\cite{TokenLake} introduces a unified segment-level prefix cache pool.
Compared to them, \sysname targets storage backend directly, balancing the traffic among all SNICs, and reduces DRAM usage greatly without harming performance.
\sysname can also be combined with a middle DRAM cache, but the performance gain is marginal. 

\parabf{KV-Cache I/O Optimization.}
Efficiently loading the massive KV-Cache from other caching tiers is a fundamental bottleneck in disaggregated LLM serving architectures.
Prior work has approached this problem primarily from the perspective of a single data path.
Strata~\cite{Strata} tackles I/O bottlenecks in hierarchical storage by co-designing GPU-assisted I/O with cache-aware scheduling.
Other work, like KVPR~\cite{ACL25:KVPR} and TailorKV~\cite{ACL25:TailorKV}, mitigates bandwidth constraints (e.g., PCIe) on this path via recomputation overlapping and layer-granular hybrid quantization. 

\parabf{LLM Inference System.}
Recent years have seen many inference acceleration techniques, such as paged attention \cite{SOSP23:PagedAttention}, chunked prefill, and hybrid batching \cite{OSDI24:SarathiServe,FastGen}. Prefill-decode disaggregated inference \cite{ISCA24:Splitwise,OSDI24:DistServe} separates the prefill and decode stages onto different GPUs, reducing performance interference between them and allowing each stage to adopt distinct parallel strategies and hardware configurations, which unlocks substantial optimization opportunities. It has largely become the de facto standard for inference serving.

\parabf{Attention Mechanisms.}
Attention mechanism allows tokens to interact with previous tokens in the sequence. 
There are many variants such as Multi-Head Attention (MHA) \cite{NIPS17:Transformer}, Multi-Query Attention (MQA) \cite{MQA} and Grouped-Query Attention (GQA) \cite{EMNLP23:GQA}, Multi-head Latent Attention (MLA) \cite{DeepSeekV2}.
For those attention mechanisms (denoted as \emph{dense attention}), the ratio of computation and KV-Cache size for one token is a constant since both scale linearly with sequence length.

\section{Conclusion}
\label{sec:conclusion}

This paper presents \sysname, an agentic LLM inference framework that addresses the imbalance of KV-Cache reading under PD-disaggregated architecture through dual-path KV-Cache loading. By redistributing storage network load with workload-aware scheduling, \sysname achieves up to 1.87$\times$ throughput improvement for offline inference. It also achieves 1.96$\times$ higher agent runs per second on average in online serving.

% % \textbf{Acknowledgement.} 
% \begin{acks}
% We use generative AI for grammar checks and language polishing. This work does not raise any ethical issues.
% \end{acks}

\bibliographystyle{ACM-Reference-Format}
\bibliography{reference}

\appendix
\section{Appendix}

\subsection{Traffic Isolation Configuration Details}
\label{sec:traffic-isolation-config}

\parabf{InfiniBand.}
The InfiniBand QoS mechanism employs two arbitrators: high-priority and low-priority.
Traffic is scheduled using Weighted Round Robin (WRR) in the high-priority arbitrator, then steered to the low-priority arbitrator according to \texttt{qos\_high\_limit}; setting it to 255 disables the low-priority arbitrator entirely.
Detailed scheduling algorithm can be found in \cite{ibta-spec}.
Our configuration:
\begin{itemize}[noitemsep]
\item \texttt{qos\_max\_vls 4}
\item \texttt{qos\_high\_limit 240}
\item \texttt{qos\_vlarb\_high 0:192,1:192,2:0,3:192}
\item \texttt{qos\_vlarb\_low 0:192,1:192,2:64,3:192}
\end{itemize}

\parabf{RoCE.}
RoCE enforces QoS through DSCP-based traffic classification and hardware traffic classes (TC). Packets are first mapped from DSCP values to TCs, each backed by a dedicated hardware queue on NICs and switches (typically up to eight). To match the four-VL configuration in InfiniBand, we configure four lossless RDMA TCs with Priority Flow Control (PFC) enabled. Bandwidth isolation is achieved by assigning proportional scheduling weights to these TCs on both NICs and switches, reserving the majority of bandwidth for model inference traffic while allocating a small fraction to KV-cache traffic to prevent starvation.

\subsection{27B Model Specifications}
\label{sec:27b-model-specs}

In terms of overall model scale, 
the hidden dimension 2560,
the intermediate size of dense layers is 12288,
the number of hidden layers is 30,
the number of attention heads is 32,
the number of routed experts is 72,
the MoE intermediate size is 1536,
the number of activated experts per token is 6, 
the number of shared experts is 2, 
and the number of initial dense layer is 1.
Regarding the index attention mechanism, 
the number of attention heads is 32,
the head dimension is 64, 
the topk tokens for sparse attention is 1024.
The LoRA compression for the Q matrix of both indexer and main attention is removed.

\subsection{Agent Task Structure}
\label{sec:agent-task-structure}

To provide context for the dataset characteristics, we briefly describe the agent task structure, though this background is orthogonal to our system design.
Each agent operates within a sandbox environment containing a code repository with known bugs and associated error messages.
The agent is instructed via prompt to diagnose and fix the bug.
The model possesses tool-use capabilities, invoking bash commands in the sandbox by emitting structured outputs.
The agent and environment engage in multi-turn interactions, where each turn consists of a prompt (previous context concatenated with new information, most of which is tool output) and the model generating a subsequent tool invocation by decoding.

Each trajectory is a sequence of rounds; round $i$ consists of appended tokens $A_i$ and the number of generated tokens $g_i$.
We use $G_i$ to indicate the tokens generated in round $i$, which are not presented in our dataset.
We define $Context_{i+1}$ as the concatenated list of $A_1, G_1, A_2, G_2, ..., A_i, G_i$.
In round $i+1$ of our replay, the agent concatenates the prompt as $Context_{i+1} + A_{i+1}$, and then sets proper sampling parameters to ensure it generates exactly $g_{i+1}$ tokens, i.e., $G_{i+1}$.
To generate additional agent trajectories, we sample an existing trajectory and prepend a synthetic round with random tokens as $A_1$ and $g_1=1$.

\subsection{Experimental Configurations}
\label{sec:experimental-configurations}

\textbf{Configuration Parameters.}
For DeepSeek models, \sysname allocates 80GB DRAM each node, and SGL(MC) uses totally 1.5TB DRAM on every node.
For Qwen 32B, due to the larger KV-Cache, \sysname allocates 320GB.
Speculative decoding is disabled for all settings.
We use 3FS as the storage backend for all configurations.
The short reading queue threshold $\alpha$ (described in \autoref{sec:scheduling}) is set to the number of tokens we can read during $3$ seconds, and the unfinished token upper limit $\beta$ is set to the number of tokens one GPU can process for $5$ seconds. Those values are profiled in advance.
Compute Quota Threshold is set to 300ms for all \sysname and Oracle baselines.

\textbf{KV-Cache Hit Length Calculation.}
For all systems except SGL(MC), we limit the KV-Cache hits to only occur within a trajectory, and the hit length is calculated in the client because no eviction is needed.
For SGL(MC), hit lengths are computed internally based on HiCache and Mooncake Store cache states.

\subsection{KV-Cache Block Layout}
\label{subsec:block-layout}
Layerwise prefill reduces KV-Cache block size to $1/layer$ of the original, and makes the number of blocks larger to $layer\times$, posing challenges to transfer and storage performance.
To overcome this, we design two distinct block types: \emph{Layer Block} and \emph{Full Block}.
A Layer Block is a byte tensor with shape $[1, tokens, bytes]$ and stores one-layer KV-Cache for some tokens.
The number of tokens is called $block\_size$.
$bytes$ indicates the cache bytes needed per layer per token.
Meanwhile, a Full Block has shape $[layer, tokens, bytes]$.
This design enables us to avoid manual KV-Cache memory layout conversion throughout inference by simply concatenating $n$ Layer Blocks to yield a Full Block.
KV-Cache is stored in distributed storage using a trie structure, where each tree node corresponds to a Full Block.

\end{document}